\documentstyle[11pt,paspconf,epsf]{article}

\begin{document}

\title{High Signal-to-Noise GHRS Observations of H 1821+643: 
       O VI Associated with a Group of Galaxies at z = 0.226 and Complex
       Lyman $\alpha$ Absorption Profiles at Low Redshift}

\author{Todd M. Tripp,\altaffilmark{1,2} Limin Lu,\altaffilmark{3,4} and Blair 
D. Savage\altaffilmark{1}}
\altaffiltext{1}{Department of Astronomy, University of Wisconsin, Madison, WI 53706}
\altaffiltext{2}{Present address: Princeton University Observatory, Princeton, 
NJ 08544}
\altaffiltext{3}{Department of Astronomy, California Institute of Technology, Pasadena, CA 91125}
\altaffiltext{4}{Hubble Fellow}

\setcounter{footnote}{3}
\begin{abstract}
As part of a program to study the relationship between QSO Lyman $\alpha$
clouds and galaxies at low redshifts, we have obtained a high signal-to-noise
(S/N) far UV spectrum of H 1821+643 ($z_{\rm em}$ = 0.297) with the G140L
grating of the GHRS.  The spectrum, which has a resolution of FWHM $\approx$ 150
km s$^{-1}$ and S/N $\approx$ 100:1, is adequate for detection of {\it weak}
Ly$\alpha$ clouds with equivalent width as low as $W_{\lambda} \ \approx$
50 m\AA\ (4$\sigma$) with complete spectral coverage from $z_{\rm abs}$ = 
0.03 to $z_{\rm abs}$ = 0.26. In this paper we present some preliminary 
results of this study including the following. (1) The absorption profiles of 
three out of the four strongest extragalactic Ly$\alpha$ lines show 
complex component structure with a main strong component and several weaker 
outlying components spanning a full velocity range of $\sim$1000-1500 
km s$^{-1}$. The two-point correlation function does not show any evidence of 
Ly$\alpha$ cloud clustering but suffers from small number statistics. 
(2) Extragalactic {\sc O vi} is clearly detected in the intervening absorption 
system at $z_{\rm abs}$ = 0.225.  Two galaxies at $z$ = 0.2256 and $z$ = 
0.2263 are close to the sight line (the closer galaxy is at a projected 
distance\footnote{Throughout this paper we use $H_{0}$ = 100 km s$^{-1}$ Mpc$^{-1}$.} of $\sim$90 kpc), so this {\sc O vi} absorption could be due to the 
intracluster medium of a group of galaxies.
\end{abstract}

\vspace{-0.25cm}
{\footnotesize To appear in {\it The Scientific Impact of the Goddard High 
Resolution Spectrograph}, Astronomical Society of the Pacific Conference 
Series, in press.}

\section{A New GHRS Study of Ly$\alpha$ Clouds -- Motivation}

Quasar absorption lines provide a powerful tool for probing the evolution 
of the universe from $z$ = 0 to 5, but to correctly interpret this information,
one must understand the nature of the absorbers. 
Studies with the GHRS have yielded important results
on the nature of low redshift ``Ly$\alpha$ clouds,'' low column density 
gas clouds which produce {\sc H i} Ly$\alpha$ absorption lines 
in the spectra of background QSOs (see Morris 1996 for a
brief review).  GHRS observations of 3C273 shortly after the deployment of
HST revealed that there are considerably more Ly$\alpha$ clouds at low redshift
than expected based on the observed evolution of the number of clouds per
unit redshift ($dN/dz$) at high $z$ (Morris et al. 1991; see also the
FOS study of Bahcall et al. 1991). This abundance of low $z$ clouds provides an opportunity to learn
about the nature of the absorbers by {\it directly} studying the environment
(i.e., galaxies, galaxy clusters, voids, etc.) where the Ly$\alpha$ clouds 
are found.  Studies of the relationship between low $z$ Ly$\alpha$ clouds and 
galaxies 
using the GHRS have been carried out by Morris et al. (1993), Stocke et al. 
(1995), and Shull, Stocke, \& Penton (1996).  These programs find that (1)
Ly$\alpha$ clouds are not randomly distributed with respect to galaxies, but
the absorber-galaxy correlation is not as strong as the galaxy-galaxy 
correlation, and (2) some Ly$\alpha$ clouds are found in galaxy voids, although
overall the clouds tend to ``avoid the voids.'' In 
general, the Ly$\alpha$ clouds studied with GHRS do not have nearby 
associated galaxies; Morris et al. (1993) report that there are no galaxies 
observed within 230 kpc of any of the 3C273 Ly$\alpha$ clouds, and Stocke et al. find 
that there are no galaxies within 450 kpc of their Ly$\alpha$ absorbers. These 
GHRS results seem to be in conflict with HST FOS studies 
of Ly$\alpha$ clouds.  For example, based on a redshift survey of galaxies 
near QSOs observed with the FOS, Lanzetta et al. (1995) find that 32-60\% of 
the Ly$\alpha$ clouds in their sample are associated with luminous galaxies 
within $\sim$ 160 kpc of the QSO sight lines. To reconcile these discordant
results, it has been 
suggested that there are two populations of Ly$\alpha$ clouds at low 
redshift: (1) strong Ly$\alpha$ clouds with $N$({\sc H i}) $\geq \ 10^{14}$ 
cm$^{-2}$ which dominate the Ly$\alpha$ cloud sample of Lanzetta et al. and 
mostly occur in large halos of luminous galaxies, and (2) lower column density 
absorbers which are less closely tied to galaxies and are, in some cases, 
truly intergalactic gas clouds  
(the Morris et al. and Stocke et al. Ly$\alpha$ clouds have 
$N$({\sc H i}) $\leq \ 5\times 10^{13}$ cm$^{-2}$.) Currently this suggestion
cannot be rigorously tested, however, because the sample of weaker Ly$\alpha$
clouds is small.

\begin{figure}
\plotfiddle{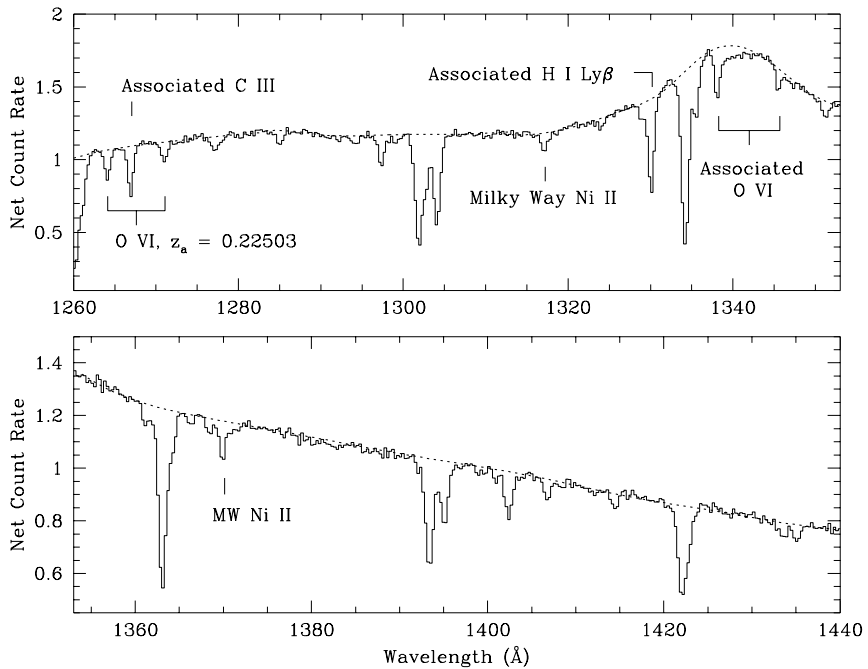}{3.0in}{0}{60}{60}{-180}{-135}
\caption{A portion of the high S/N spectrum of the QSO H1821+643 obtained
with the GHRS using the G140L grating.}
\end{figure}

To significantly improve the statistics of low redshift {\it weak} Ly$\alpha$
clouds, we are conducting a program to study the relationship between low $z$
Ly$\alpha$ clouds and galaxies in the direction of three QSOs using the GHRS 
and the 
WIYN multiobject spectrograph (HYDRA). Using the G140L grating, we will obtain
GHRS spectra with S/N $\approx$ 100:1, adequate for detection of Ly$\alpha$ 
clouds with $W_{\lambda}$ = 50 m\AA\ (4$\sigma$), and this will increase the 
{\it weak} Ly$\alpha$ cloud sample size by a factor of 4-5. The WIYN HYDRA 
will be used to measure the redshifts of galaxies in the $\sim 1^{\circ}$ fields
centered on the QSOs. Figure 1 shows a portion of the GHRS spectrum of 
H 1821+643 ($z_{\rm em}$ = 0.297) obtained for this program. The WIYN galaxy
redshift survey for this field has been completed, and the full analysis of
the GHRS and WIYN data for this sight line will be presented in a subsequent
paper.  Some preliminary results are summarized below.

\section{Highly Ionized Oxygen at {\it z} = 0.225}

The {\sc O vi} 1031.9, 1037.6 \AA\ doublet is well-detected at $z_{\rm abs}$
= 0.2250 in the H 1821+643 spectrum (see Figure 1) along with {\sc H i} 
Ly$\alpha$ and Ly$\beta$, and we
have G160M (FWHM = 15 km s$^{-1}$) GHRS observations of the
{\sc O vi} 1031.9 \AA\ and {\sc H i} Ly$\beta$ absorption profiles 
obtained in the Galactic ISM program of Savage et al. (1995).
These G160M profiles are well-described by Voigt profiles with the parameters
listed in Table 1. The width of the {\sc O vi} profile indicates that 
$T \ \leq \ 1.8\times 10^{6}$K, and the redshift difference between the 
{\sc H i} and {\sc O vi} profiles implies a 40 km s$^{-1}$ centroid shift 
between the neutral and highly ionized gas. The {\sc H i} Ly$\alpha$ profile
shows complex component structure (see Figure 2), and some of this structure
could be due to a broad hot {\sc H i} component associated with the {\sc O vi}.
\begin{table}
\caption{O VI Absorber -- Profile Fitting Results}
\begin{tabular}{lccc}
\tableline
Species & Redshift              & log $N$ (cm$^{-2}$) & $b$ (km s$^{-1}$) \\ \tableline
H I Ly$\beta$ & 0.224892$\pm$0.000008 & 15.32$\pm$0.07      & 50.7$\pm$3.2 \\
O VI          & 0.225026$\pm$0.000010 & 14.29$\pm$0.03      & 42.8$\pm$3.5 \\
\tableline \tableline
\end{tabular}
\end{table}

Schneider et al. (1992) have detected an emission line galaxy at $z$ = 0.2256
within 90 kpc of the sight line. Our WIYN redshift survey has discovered
another emission line galaxy close to the sight line at $z$ = 0.2263. 
Therefore it is possible that the {\sc O vi} absorption at $z$ = 0.2250 
originates in the intracluster medium of a group of galaxies.

\section{Clustered Ly$\alpha$ Clouds?}

The absorption profiles of three out of the four strongest {\sc H i} Ly$\alpha$
lines contain complex component structure with a main strong component and 
several weaker outlying components spanning $\sim$1000-1500 km s$^{-1}$ (see 
Figure 2). Even the Ly$\alpha$ line at $z_{\rm abs}$ = 0.1699 which does not 
show resolved weak components has an asymmetric profile which is evidence 
of unresolved profile components.  This seems to suggest that there is some
clustering of weak Ly$\alpha$ clouds at low redshift. However, 
we have calculated the two-point
correlation function using all of the Ly$\alpha$ clouds detected in the
H 1821+643 GHRS spectrum, and this does not show evidence of clustering on
any scale, but the sample of Ly$\alpha$ lines may be too small for adequate
statistics.
\begin{figure}
\plotone{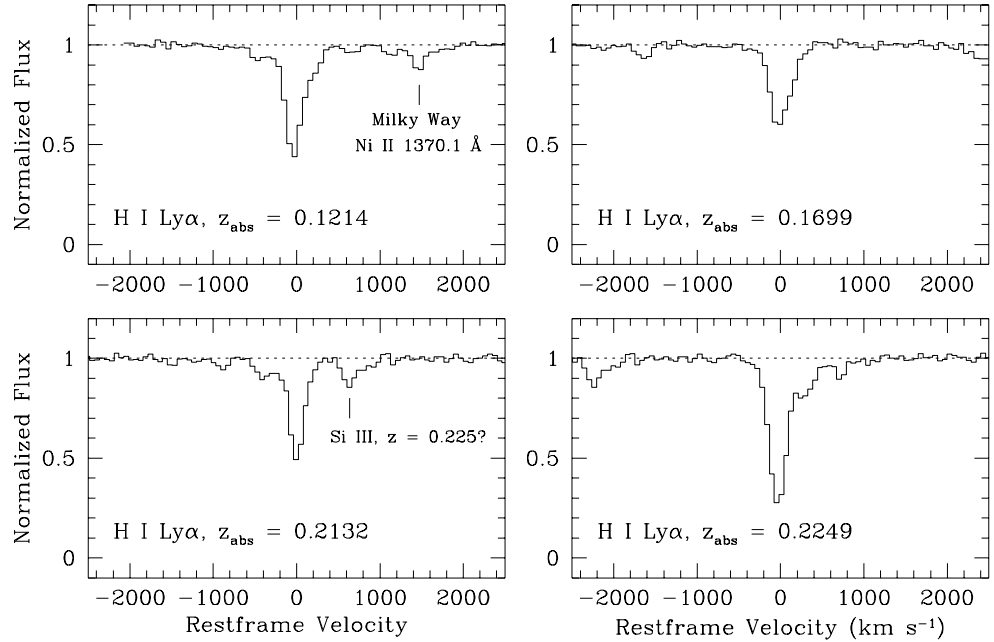}
\caption{Absorption profiles of the four strongest {\sc H i} Ly$\alpha$ lines
detected in the GHRS spectrum of H 1821+643.}
\end{figure}

\acknowledgements Our WIYN galaxy redshift program is a collaboration with 
Buell Jannuzi. This work is supported by NASA through grant NAG5-1852.

\end{document}